\documentclass[fleqn,10pt]{wlscirep}
\usepackage[T1]{fontenc}

\usepackage{bm}
\usepackage{rotating}
\usepackage{subfigure}
\usepackage{multirow}
\usepackage{amsmath,amssymb}

\title{
Differences of communication activity and mobility patterns between urban and rural people
}

\author[1,*]{Fumiko Ogushi}
\author[2]{Chandreyee Roy}
\author[2,3,$\dagger$]{Kimmo Kaski}

\affil[1]{Center for Mathematical Modeling and Data Science, Osaka University}
\affil[2]{Department of Computer Science, Aalto University School of Science, Espoo, Finland}
\affil[3]{The Alan Turing Institute, London, UK}

\affil[*]{ogushi@sigmath.es.osaka-u.ac.jp}
\affil[$\dagger$]{kimmo.kaski@gmail.com}


\begin{abstract}
Human mobility and other social activity patterns influence various aspects of society such as urban planning, traffic predictions, crisis resilience, and epidemic prevention. The behaviour of individuals, like their communication frequencies and movements, are shaped by societal and socio-economic factors. In addition, the differences in the geolocation of people as well as their gender and age cast effects on their activity patterns. In this study we focus on investigating these patterns by using mobile phone data, specifically the call detail records (CDRs), to analyze the social communication and mobility patterns of people. This dataset can provide us insight into the individual and population-level behaviours in rural and urban environments on a daily, weekly and seasonal basis.  
The results of our analyses show that in the urban areas people have high calling activity but low mobility, while in the rural areas they show the opposite behaviour, i.e. low calling activity combined with high mobility. Overall, there is a decreasing trend in people's mobility through the year even though their calling activity remained consistent except for the holidays during which time the communication frequency drops markedly. We have also observed that there are significant differences in the mobility between the work days and free days. Finally, the age and gender of individuals have also been observed to play a role in the seasonal patterns differently in urban and rural areas. 
\end{abstract}
\begin{document}

\flushbottom
\maketitle

\thispagestyle{empty}

\section*{Introduction}
Humans, being creatures of habit, inherently reproduce behavioural and moving patterns over periods of time even though their daily activities may be diverse \cite{gonzalez2008understanding}. Several phenomena including urban planning \cite{horner2001embedding,hillier2009metric,wu2020comparison}, traffic predictions \cite{kitamura2000micro,de2011modelling}, crises resilience 
\cite{lu2016detecting,lu2012predictability} and prevention of epidemics \cite{kraemer2020effect, balcan2009multiscale} among others are influenced by the habitual patterns of humans. 
Specifically their social activity and mobility behaviour, such as how frequently and in what ways they communicate with others, as well as how actively they move locally or travel more distantly, are influenced by the society and environment in which they live and reflect the socio-economic variables of their lives \cite{kraemer2020mapping}. Consequently, differences in living and social conditions of different geographical locations can affect these patterns \cite{meredith2021characterizing}. For example, some studies indicate that individuals in rural areas have higher mobility than in urban areas \cite{pucher2004urban} and it has also been observed that gender \cite{gauvin2020gender} and age \cite{gimenez2022daily} play a role \cite{liu2023uncovering}. 

The patterns of human mobility and activity have implications in our lives\cite{gilibert2019analysis}, politics \cite{wicki2019does} and sustainability of environment \cite{pyky2019individual}. The daily behaviour of individuals \cite{ferrer2018associations} are governed by personal choices \cite{borst2009influence, guo2013pedestrian}, which have implications at the social and environmental levels \cite{machado2018overview, mourad2019survey}. 
Our aim in this study is to understand and get insight into potentially universal aspects of communication activity and human mobility patterns and the important characteristic variables leading to them.  
In the literature, there are several studies that have developed predictive models for human mobility \cite{spadon2019reconstructing, liu2020universal, smolak2021impact} to assist for better urban planning \cite{louail2015uncovering} or to help policy makers create more effective plans in case of epidemic spreading \cite{hufnagel2004forecast, eubank2004modelling, balcan2009multiscale} or other crisis situations \cite{lammel2010representation, kunwar2016evacuation}. 

As the use of mobile phones is so prevalent in the everyday lives of the society, the digital data generated by the people using them offers us a quite unprecedented view to their social communication activities \cite{eagle2009inferring, monsivais2017tracking,jo2012circadian,onnela2007structure} and mobility patterns \cite{astero2022mobility}, on the daily, weekly and seasonal basis at the individual as well as population level \cite{bhattacharya2019social}. 
In this way generated datasets consist of call detail records (CDRs) of anonymized individuals with communication activity (call or text message) recorded in the data and associated with a cell tower in the vicinity of the individual making the call. 
This gives us information about the approximate location of the individual 
at a certain date and time during the activity. Such location information during the course of the day allows us to trace the trajectories of individuals and to identify regular patterns of mobility in the data \cite{fudolig2021internal,astero2022mobility}. These kinds of datasets are particularly useful in studying human behaviour in the large scale emergency situations \cite{wang2020using}. 

In the present study, we focus on a large scale dataset of mobile phone usage from a Southern European country. We investigate the daily, weekly and seasonal communication activity and mobility patterns of the mobile phone service subscribers and their behavioural differences in the urban and rural areas along with their age and gender differences. 
Through the daily patterns of calling activity we determine the ``home'' location for each individual to be the night time period of low communication activity. We also identify the distribution of population density in the country based on the number density of the service subscribers in a given area with reference to their ``home'' location. Our analyses indicate that while the urban and rural areas have different population densities, they can also be distinguished by the service subscribers' weekly and seasonal behaviour patterns. 
The calling activity is observed to be high for the individuals in the urban areas while they have lower mobility in comparison to rural areas. 
We find the average mobility to be different on work days from that of free days and there is an overall decreasing trend in the mobility of people towards the end of the year. Additionally, the age and gender also turn out to play a role by affecting their seasonal patterns of the calling activity and mobility.

The paper is organized such that in the next section 2 we identify regions in the country that are densely populated and compare their communication activity and mobility patterns with low population density regions. 
In section 3, we focus on studying these differences in the regions with respect to their weekly and seasonal  behaviour and finally in Section 4, we finish with a discussion of our results. 

\section{Basic patterns of communication activity and mobility}
\subsection{Daily pattern and ``home'' location}
\begin{figure}[h!]
\centering
\includegraphics[width=0.6\linewidth]{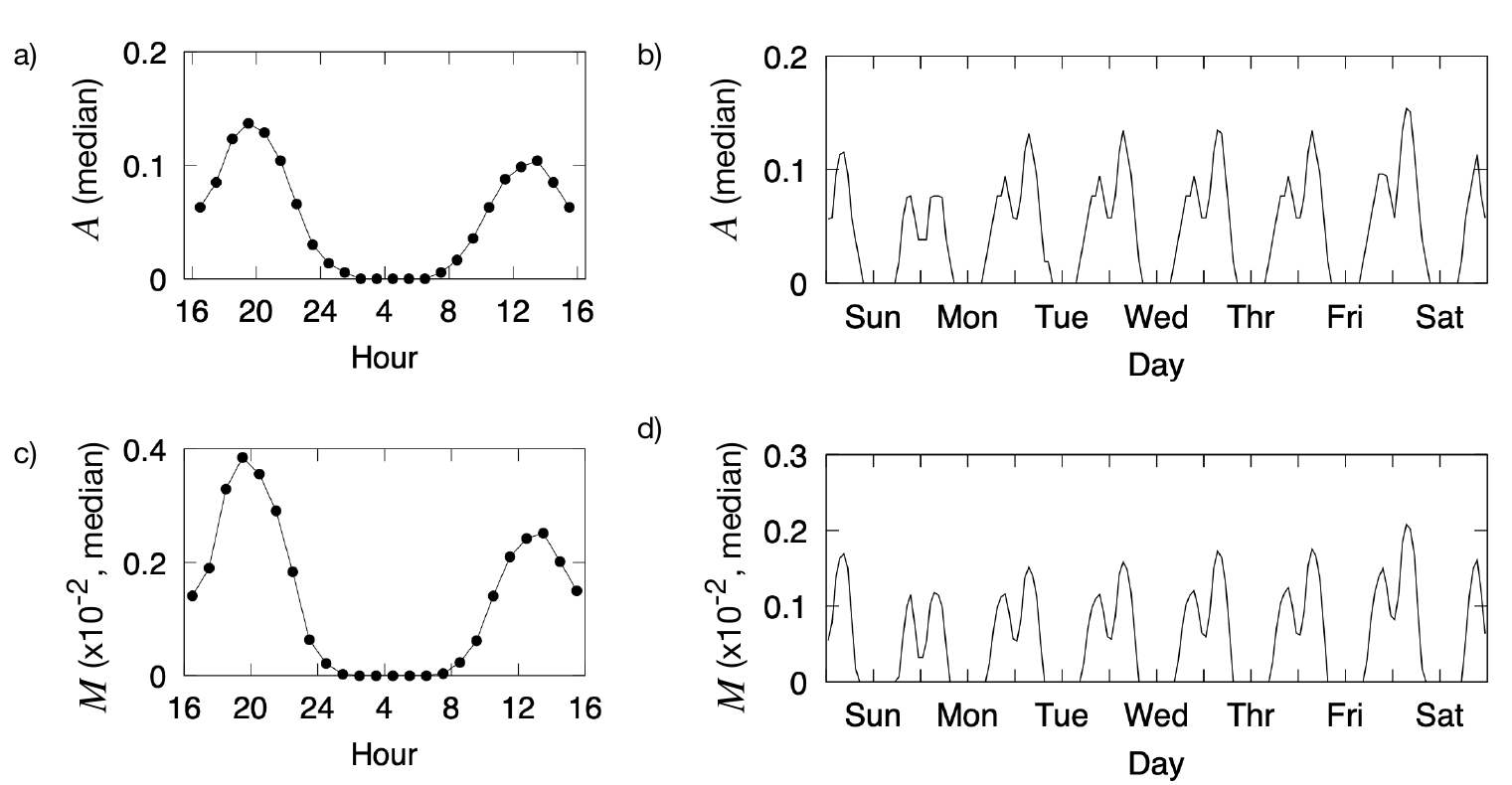}
  \caption{
    Daily patterns of human communication activity and mobility.
    Average patterns of daily communication activity (a) and mobility (c) of all individuals. 
    Weekly variation of daily pattern of communication activity (b) and mobility (d). 
    Both the communication activity and mobility show a bimodal distribution with peak around 1 pm and 8pm and both zero around 5 am, i.e. $A=M=0$. The two peaks of the communication activity are well fitted with the Gaussian distribution function $G(\mu, \sigma)$ where the day-time peak is at ($\mu_{day}=$12:58 pm, $\sigma_{day}=$2.36) and the evening peak at ($\mu_{night}=$7:43 pm, $\sigma_{night}=$2.31). The low activity night time is defined to take place from 1 am to 7 am on a basis of being the most inactive period.
    }
    \label{fig_daily_pattern}
\end{figure}

Let us start by calculating the communication activity $A_i$, the mobility $M_i$, and the ``home'' location $\bm{r}_i^{\rm{home}}$ of each individual $i$ from the CDR data. The daily patterns of the communication activity and mobility of individuals show a bimodal distribution with peaks around 1 pm and 8 pm along with their lowest values occurring at around 5 am, as depicted in Fig.~\ref{fig_daily_pattern}(a) and (c), respectively. 
The bimodal distributions of both the communication activity $A$ and mobility $M$ remain the same through the week see Fig.~\ref{fig_daily_pattern}(b) and (d)). 
In the case of communication activity, both peaks can be well fitted with the Gaussian distribution functions. The typical time scale of communication on mobile phone is shorter than that of the actual movement of individual, suggesting that the communication activity can reflect a more detailed schedule of our social life. 
In fact, compared to the mobility, the daily pattern of communication activity has a broader and flatter bottom as well as narrower peaks. 

By assuming that people stay at home during the night time period of low activity, we define the ``home'' location of an individual $i$ by taking the average of all tower locations for calls made during this time period. Based on the daily patterns of the communication activity and mobility, the night time period is determined to be in between 1 am and 7 am. The daily pattern of radius of gyration from the ``home'' location $R_g$ also show a bimodal distribution and $R_g$ becomes almost $0$ in the night time period from 1 am to 7 am (see Fig.~\ref{sfig_average_pattern} in Appendix). 
In the following, we use the ``home'' locations to calculate the density distribution of individuals and utilize it to discuss the density dependence of the communication and mobility activity. See the Appendix for more details of the mobile phone data, and how the communication activity $A_i$, mobility $M_i$, and ``home'' location are calculated.

%%%%% daily pattern of activity and mobility

\subsection{Density dependence on communication activity and human mobility}
\begin{figure}[b!]
\centering
\includegraphics[width=0.6\linewidth]{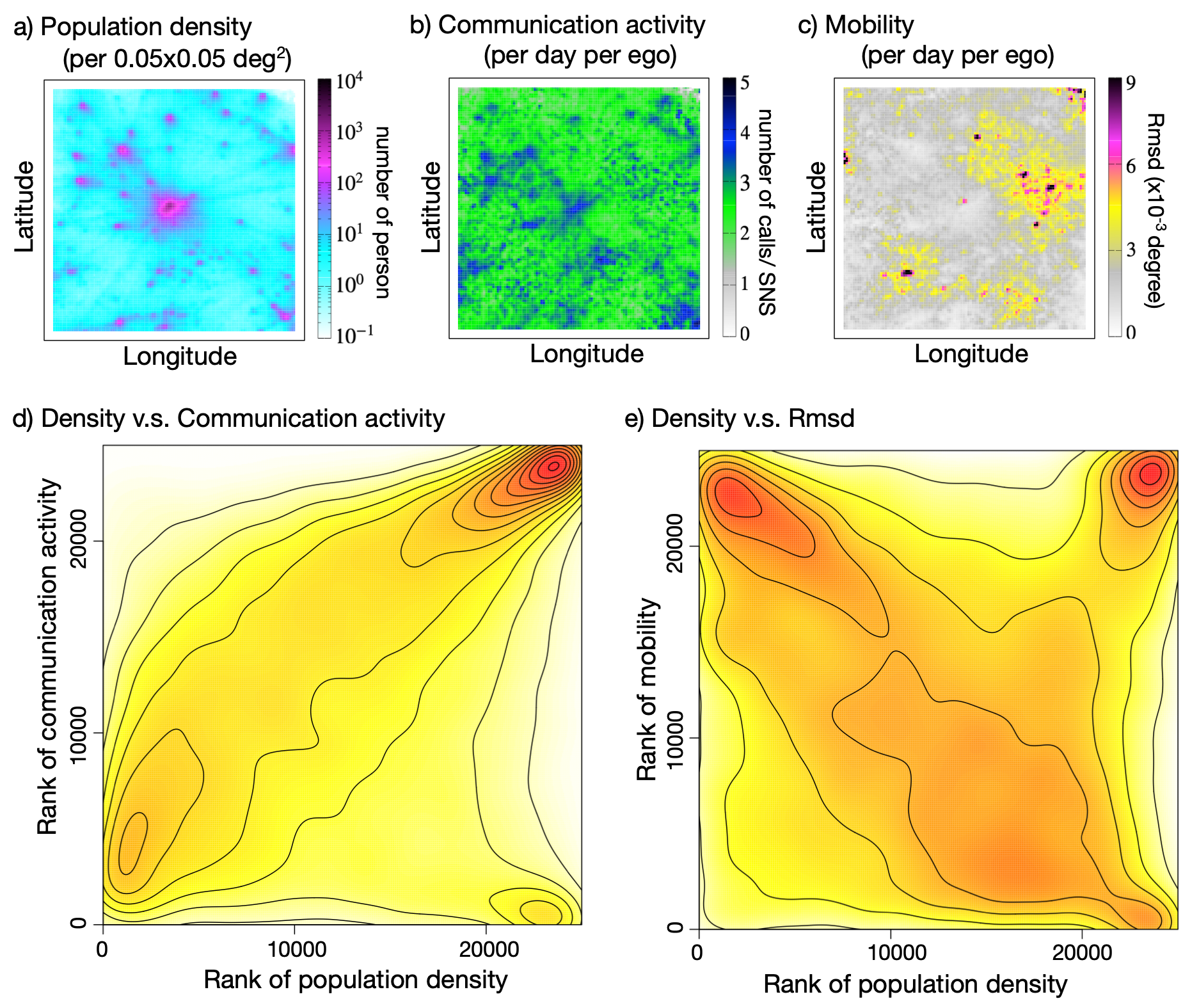}
  \caption{
    Dependence of human communication activity and mobility on population density of an area. 
    (a) Sample map of the density distribution of individuals around a typical area, (b) communication activity, and (c) mobility in that area, (d) rank-rank plot of population density and the communication activity of individuals, and (e) rank-rank plot of population density and the mobility of individuals. The communication activity of individuals correlates positively ($corr(R_{\rho}, R_A)=0.38$) while their mobility correlates negatively ($corr(R_{\rho}, R_M )=-0.11$) with the population density of individuals. 
    }
    \label{fig_densmap}
\end{figure}

In Fig. \ref{fig_densmap} we show the distributions of communication activity and human mobility in relation to the density distribution of individuals as determined by the local number density of persons in each cell of lattice with the resolution of $0.05$ latitudinal degrees times $0.05$ longitudinal degrees based on the geographical location of their ``home''. This corresponds roughly to a lattice with cell resolution of $5$ km$^{2}$. The distribution of human communities with various densities and sizes in a certain geographic location of the country is displayed in Fig.~\ref{fig_densmap}(a) and it can be used to identify regions of different population densities and area sizes, i.e. cities, towns and villages. In Fig.~\ref{fig_densmap}(b) and (c) we depict distributions of the communication activity and human mobility in the same geographical section of the country that was used to create Fig.~\ref{fig_densmap}(a). 

Next, we examine the density dependence of the communication activity and mobility of individuals. 
We find that the communication activity tends to be higher in more densely populated areas, while the mobility is negatively correlated with the density as can be seen in Fig.~\ref{fig_densmap}(d)~\ref{fig_densmap}(e), respectively. 
The third extra ``islands'' of the low density areas found in the high communication activity (bottom right of Fig.~\ref{fig_densmap}(d)) and low mobility areas (top right of Fig.~\ref{fig_densmap}(e)) are due to individuals whose ``home'' was determined to be located at sea and is an artifact of the definition of ``home'' location. 

%%%%% density dependence of A & M 
 
\begin{figure}[t!]
\centering
\includegraphics[width=0.7\linewidth]{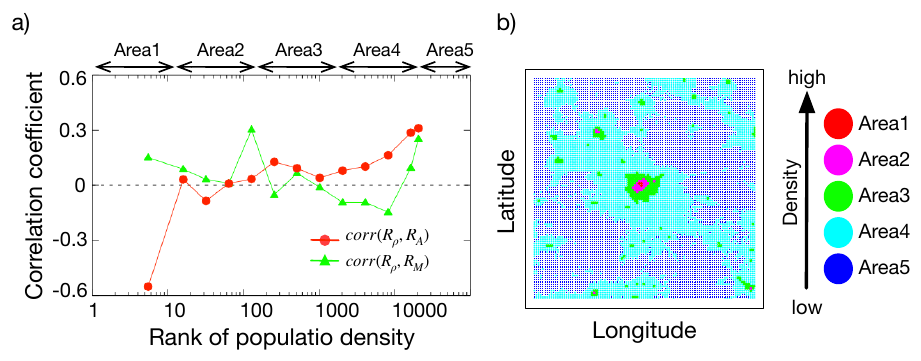}
  \caption{
    Correlation of the communication activity and mobility as a function of population density.
    (a) Correlation patterns of the communication activity and mobility changing with the local population density. 
    (b) Sample map of five areas. The population density is seen to correlate positively with the communication activity and negatively with the mobility in the densely populated areas (density rank $R_\rho < 100$). 
    In the moderate or lower population density areas, the communication activity correlates negatively and the mobility positively with the population density. For more detailed quantification the country 
    is divided into five types of areas, Area 1 to Area 5, i.e. from high to low density. 
    }
    \label{fig_corr}
\end{figure}

In order to discuss the difference of human activity between urban and rural areas, we need to clarify the basic changes of the communication activity and human mobility depending on their population densities. 
For this purpose, we investigate the correlation patterns that vary with the density of individuals using Spearman's correlation coefficient. Here the ranks have been assigned in descending order. 
In Fig.~\ref{fig_corr}(a) we show the correlation patterns between the density rank and the communication activity rank, $corr(R_\rho, R_A)$, between the density rank and the mobility rank, $corr(R_\rho, R_M)$. 

In the densely populated areas (density rank $R_\rho \leq 100$), the communication activity is negatively correlated with the population density of the location and changes to positive in the moderate or lower density areas. 
On the other hand, the correlation between the population density and the mobility, $corr(R_\rho, R_M)$, changes from positive to negative as the population density decreases. 
These results mean that as the population density increases, individuals become less mobile and communicate more frequently using their mobile phones. 
Note that the mobility is positively correlated with the density in extremely sparse areas ($R_\rho > 20,000$). This is because some of the individuals in these areas are traveling over longer distances, such as cross-country or between the mainland and an island. 
Based on the correlation patterns discussed above, 
we use the population density to divide the country into five types of areas, i.e. Area 1 to Area 5, ranging from high to low population density (see Fig.~\ref{fig_corr}(b)). The rank size distribution of population density calculated based on the subscriber's ``home'' location also suggests that the country consists of multiple high density centers ($R_{\rho} < 100$) and tail regions that are well fitted by a power law (see Fig.~\ref{sfig_drank} in Appendix). The average density in Area 3 is comparable to the actual value of population density in the country in 2008, being $90$ inhabitants per km$^{2}$.

%%%%% correlation patterns 

\section{Urban-rural differences of communication activity and mobility}
The communication activity is found to be positively correlated with the population density in both weekly and seasonal profiles, with more activity in the densely populated areas and less activity in the sparsely populated areas. On the other hand, the mobility turnes out to be negatively correlated with the population density (see Fig.~\ref{sfig_AMpatterns_nominal} in Appendix). 
In this section, we focus on weekly and seasonal differences of the communication activity and mobility, and further discuss urban vs. rural differences in these patterns, as well as the differences due to the age and gender of individuals.

\subsection{Weekly and seasonal patterns}
\begin{figure}[h]
\centering
\includegraphics[width=0.8\linewidth]{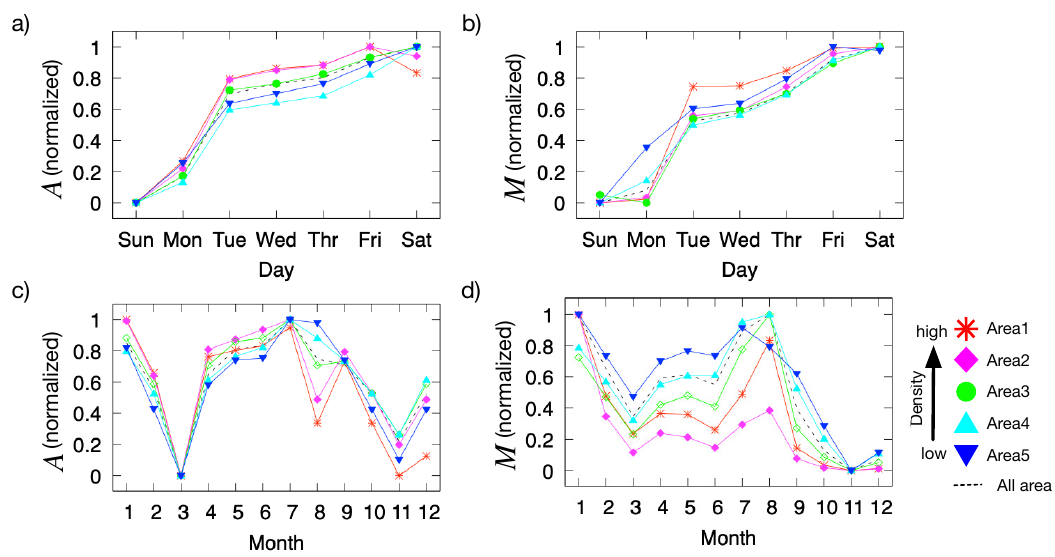}
  \caption{
    Normalized median of weekly and seasonal patterns of the communication activity and mobility by area.
    (a) Weekly pattern of communication activity. (b) Weekly pattern of mobility. (c) Seasonal pattern of communication activity. (d) Seasonal pattern of mobility.     
    Weekly patterns of communication activity are found to be similar in all five areas (Area 1 to Area 5). Individuals communicate more frequently on weekdays, with the maximum on Fridays and Saturdays  and minimum on Sundays. 
    The mobility is maximal on Fridays and Saturdays and minimum on Sundays in all areas. 
    The communication activity tend to be higher in the spring-summer season from March to August and lower in the autumn-winter season from September to December. In Area 1 and 2, the communication activity shows a clear decrease in August. 
    The mobility shows two peaks in January and July or August, with a local minimum in March and a large increase after December. 
    In the densely populated areas the mobility is seen to decrease in March more than in sparsely populated areas and after March the change in the densely populated areas is more moderate than in the sparsely populated areas.  
    }
    \label{fig_average}
\end{figure}

The weekly patterns of communication activity of people in all the five different areas show a unimodal distribution with an increase from Sunday to Friday and a decrease from Friday to Sunday as shown in Fig.~\ref{fig_average}~(a). The dominant behaviour is a fall on Sunday. 
The weekly patterns of the mobility of people also show an increase from Sunday to Friday and a peak on Friday in all the areas (Fig.~\ref{fig_average}~(b)). 

The seasonal pattern indicates that the communication between individuals tends to be more active from March to August than from September to December (Fig.~\ref{fig_average}~(c)). 
After the spring, the communication activity turns out to become more and more active peaking in July. We find that people are least active in March and make more contacts from December to January, which is consistent with the Easter holiday and the time period around Christmas and the New Year. 

The urban vs. rural differences of the communication activity and mobility of people appears mainly in the seasonal patterns. As for the area-wise behaviour patterns we find that the densely populated areas (Area 1-3) show a clear drop of communication activity in August, which is probably due to people going on holidays.
We also find that the mobility shows a decreasing trend throughout the year, with a local minimum in March and November, and then peaks at the beginning of the New Year and August (Fig.~\ref{fig_average}~(d)).
In the densely populated areas, the mobility increases almost to a maximum in January, followed by relatively speaking more moderate change from March to August than in sparsely populated areas. 
In addition, the mobility in the sparsely populated areas also reaches a maximum in January followed by a drop in March. 
In the sparsely populated areas, seasonal deviation in the mobility of people from March to November is more pronounced than in the densely populated areas (see also Fig.~\ref{sfig_AMpatterns_nominal} in Appendix). 

The above mentioned differences in seasonal and weekly patterns with respect to areas suggests that they can be grouped as follows: urban area (Area 1 and 2), rural area (Area 4 and 5), and intermediate area (Area 3).
We cross-checked the address of ``home'' locations of individuals and found that Area 1 and 2 mainly correspond to cities, Area 3 to towns, and other areas with lower density (Area 4 and 5) to villages.

%%%%% average pattern of activity and mobility 

\subsection{Gender- and age-dependence}
\begin{figure}[h!]
\centering
\includegraphics[width=0.8\linewidth]{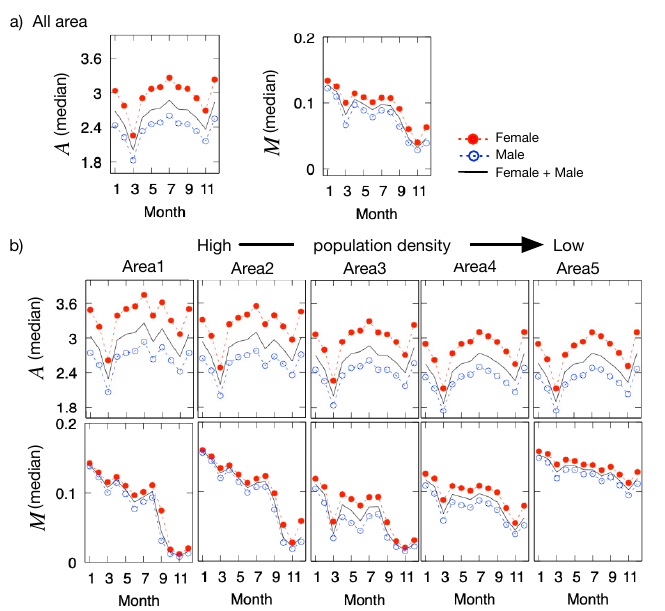}
  \caption{
    Seasonal patterns of the communication activity and mobility by the area and gender.
    (a) Seasonal patterns for all areas. (b) Seasonal patterns by area. Females (red, filled circle) show higher mobility and communication activity than males (blue, open circle) throughout the year in all areas. The communication activity is observed to be slightly higher in the densely populated areas while mobility is slightly higher in the sparsely populated areas.
    }
    \label{fig_gender}
\end{figure}

Here we focus on the gender and age differences in the communication activity and mobility of individuals through the year. 
For this purpose, we use the gender and age information of a subset of $1.8$ million service subscribers. In all areas, the females are found to be more active than males, both in terms of the communication activity and mobility, although there is no gender difference in the shapes of these patterns (Fig. ~\ref{fig_gender}). The mobility of individuals in sparsely populated areas turns out to be 
slightly higher and the seasonal variation is lower than in the densely populated areas, while the communication activity is higher in the densely populated areas. The radius of gyration from the ``home'' location shows a clear increase as the population density increases (see also Fig. ~\ref{sfig_totalAMRg} in Appendix). 
Note that in the subset data including information of gender and age, the population density dependence of the mobility is the opposite to the results observed for the entire dataset.

\begin{figure}[h!]
\centering
\includegraphics[width=0.8\linewidth]{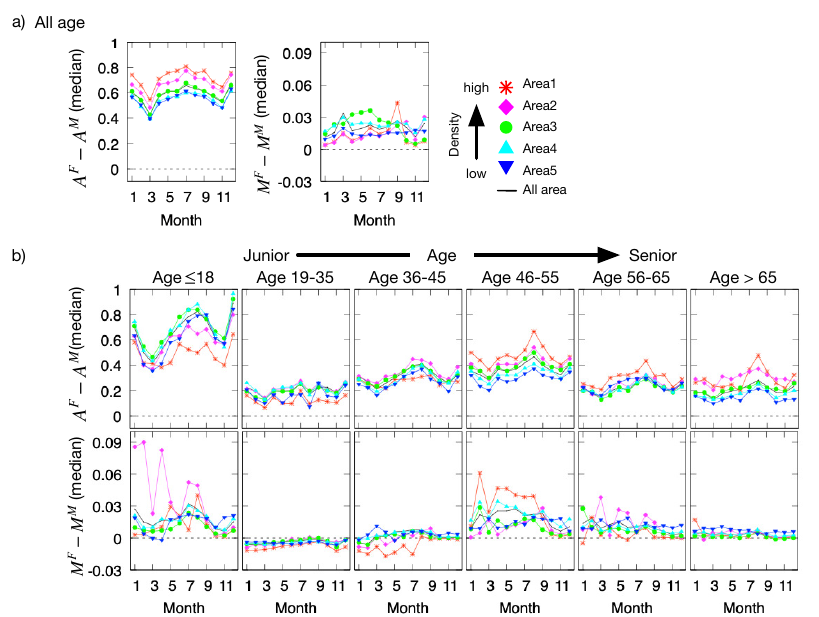}
  \caption{
    Differences of the communication activity and mobility between females and males by the age group.
     (a) Average gender differences for all age: the difference of the communication activity, ($A^{F}-A^{M}$), is positive and large and becomes larger when the population density increases, thus clearly and growingly in favour of females with the population density. The gender difference of the mobility, ($M^{F}-M^{M}$), is positive and small, thus only slightly in favour of females. 
     (b) Gender differences by the age group: the difference of communication activity in the under 18 year old teenage group is clearly in favour of females and growingly so with decreasing population density, while for the following three age groups, 19-35, 36-45, and 46-55, the difference is much smaller with slight growth by the age group after which it settles once again to a lower positive level but having slightly increasing trend with population density. 
     The differences of mobility are mostly minimal through all the age groups and for all population densities. In the age group 19-35, the gender difference of mobility tends to be slightly negative but after that it seems to turn out to be slightly positive.
    }
    \label{fig_gender_age}
\end{figure}

\begin{figure}[h!]
\centering
\includegraphics[width=0.9\linewidth]{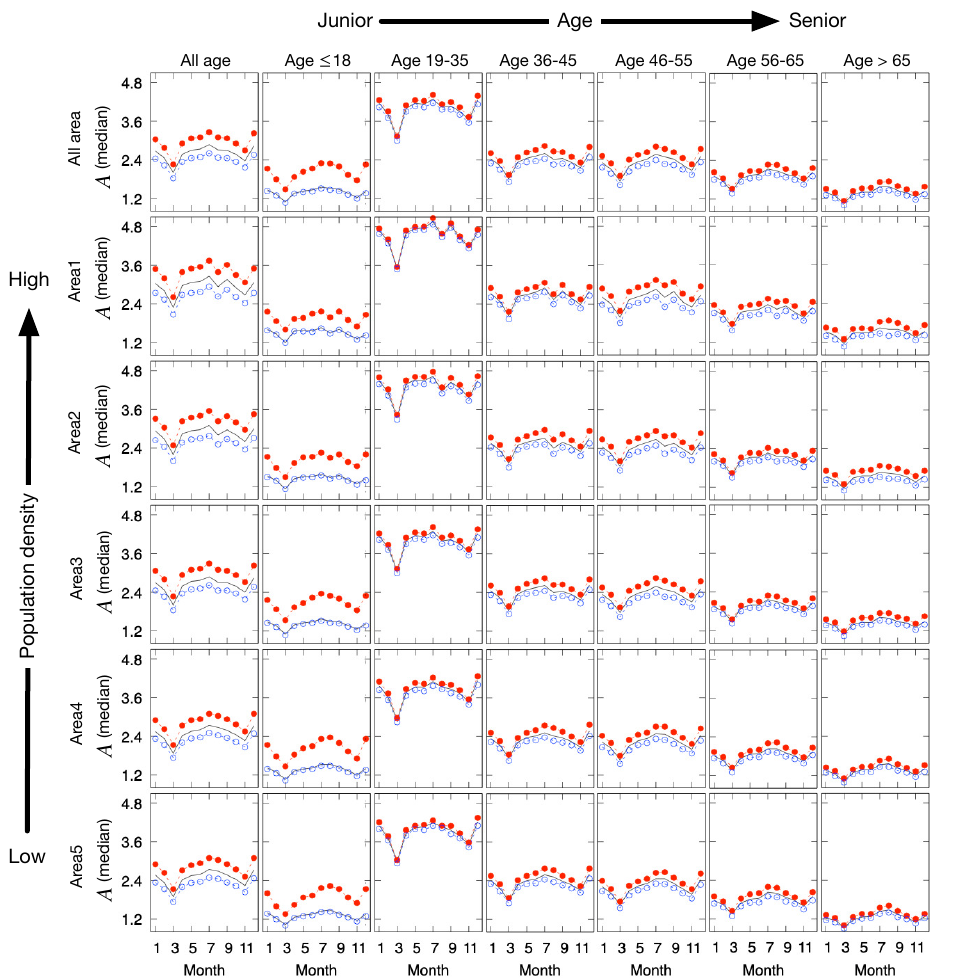}
  \caption{
    Seasonal patterns of the communication activity by the area and age group.
   Communication activity $A$ shows highest value in early adulthood in all areas and decreases with age. The gender difference of $A$ seems to be small in early adulthood, large in middle age and eventually decreasing with age after 56 year old. In the teenage group, the gender difference of $A$ is larger than other age groups. 
    }
    \label{fig_seasonal_activity_wAA}
\end{figure}

\begin{figure}[h!]
\centering
\includegraphics[width=0.9\linewidth]{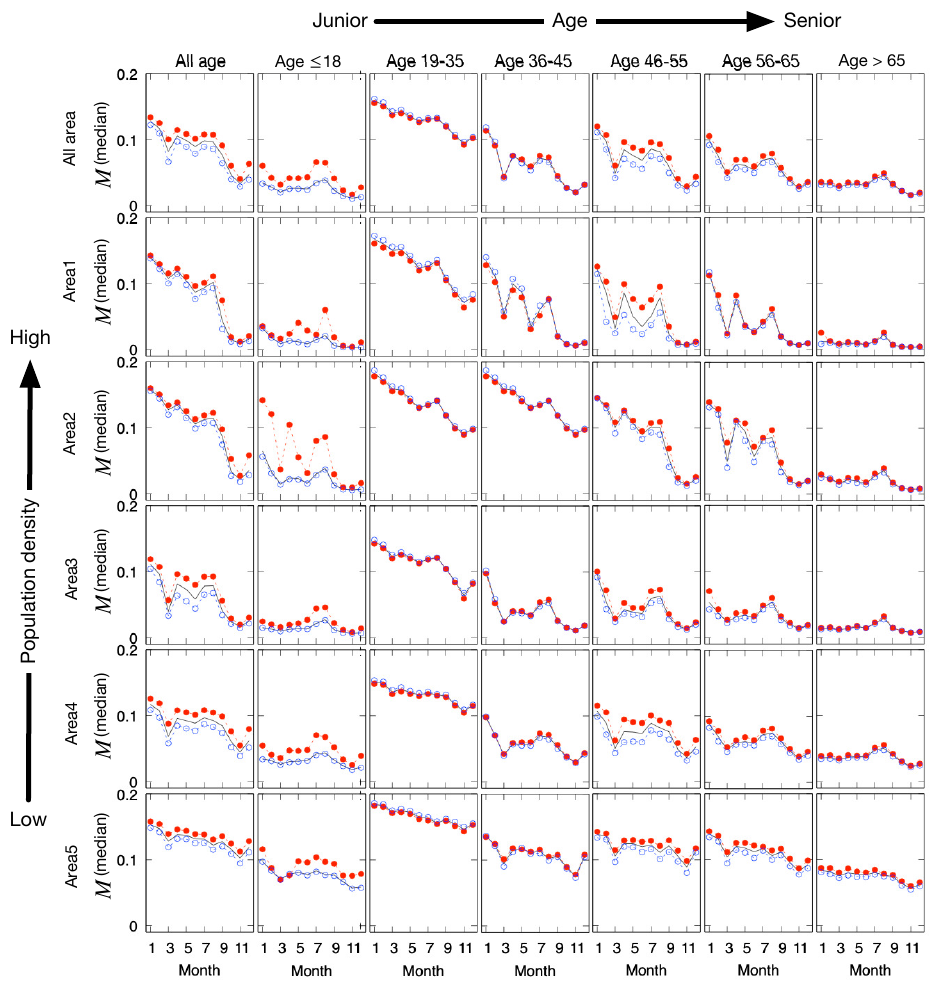}
  \caption{
    Seasonal patterns of the  mobility by the area and age group.
    Independent of age, $M$ shows larger seasonal variation in densely populated area. The gender difference of $M$ is large in teenage and middle age groups where females seem to move more actively than males. In early adulthood, males tend to have slightly larger $M$ than females.
    }
    \label{fig_seasonal_mobility_wAA}
\end{figure}

\begin{figure}[h!]
\centering
\includegraphics[width=0.9\linewidth]{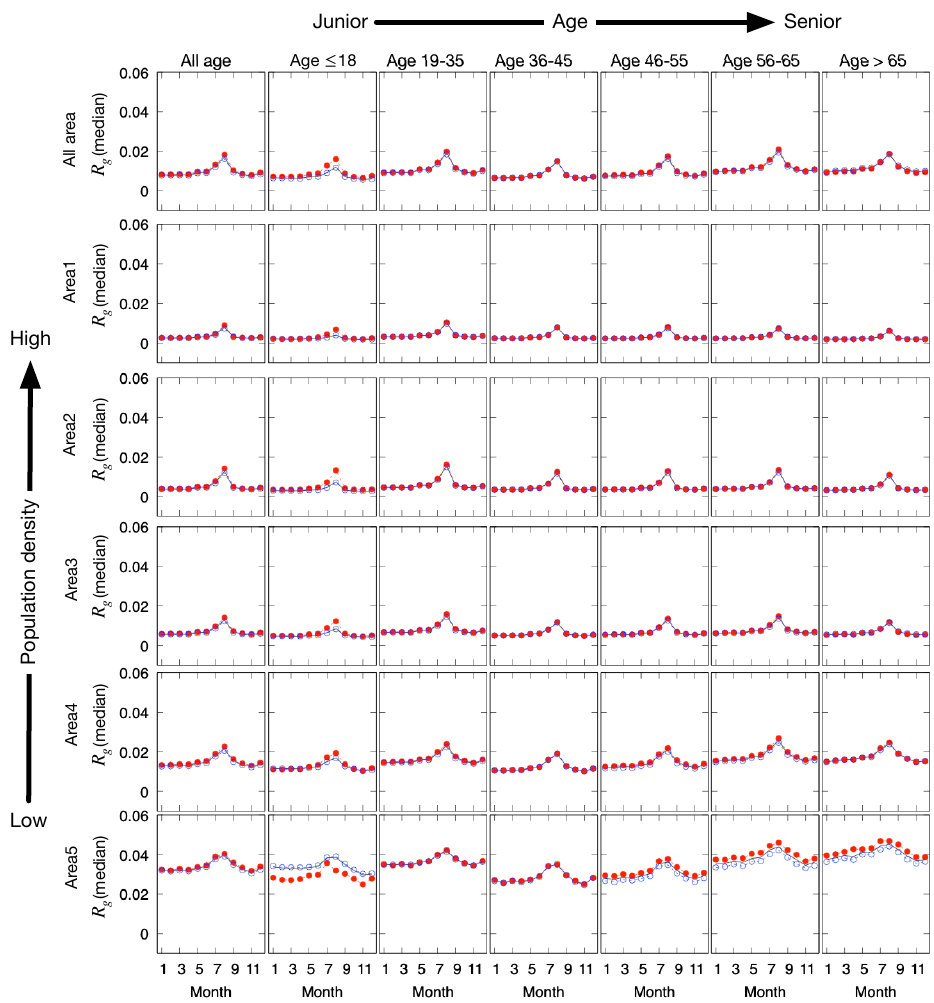}
  \caption{
    Seasonal patterns of the  radius of gyration from their ``home'' location by the area and age group.
    Sparsely populated areas have higher $R_g$ than densely populated areas and the gender difference is particularly more pronounced in the Area 5 among the teenage and middle age and older groups. All the distributions are found to be uni-model with a peak in August. 
    }
    \label{fig_seasonal_rg_wAA}
\end{figure}

In addition, the gender difference in the communication activity (i.e. female minus male communication activity) increases with the population density, while conversely, the gender difference in mobility is small and tends to be lower in the densely populated areas (Fig. ~\ref{fig_gender_age}(a)). 
As for the age differences, we divided the individuals in the following groups: teenager ($\le 18$ year old), early adulthood (between 19 and 35 year old), early middle age (between 36 and 45 year old), middle age (between 46 and 55 year old), early senior (between 56 and 65 year old), and senior ($\ge 66$ year old). 
As a result, we find the gender differences in the communication activity of people to increase with decreasing population density for the younger age group i.e. teenage, but this behaviour is reversed for people from middle aged on and showing larger differences in the urban areas (Fig. ~\ref{fig_gender_age}(b) and Fig. ~\ref{fig_seasonal_activity_wAA}). For the middle and senior age groups, the gender difference in the most densely populated areas becomes clearly larger than in the other areas. 
For early adulthood, the gender difference in the communication activity is smallest compared to other age groups while the corresponding difference in mobility is small for all age groups except for the teenage and middle age. It appears that males tend to move slightly more actively than females among the early adulthood and early middle age groups (Fig. ~\ref{fig_gender_age}(b) and Fig. ~\ref{fig_seasonal_mobility_wAA}). 
The gender and age dependence on the radius of gyration from the ``home'' location $R_g$ is depicted in Fig. ~\ref{fig_seasonal_rg_wAA} and it is found to be larger in the sparsely populated areas than densely populated ones. Additionally elder age groups have larger $R_g$ compared to younger age groups in the sparsely populated areas. In the most sparsely populated area, males seems to have larger $R_g$ than females in the teenage group but that is reversed for age groups of 46 year old and older.

%%%%% average pattern of activity and mobility

%%%%% gender difference of activity and mobility

%%%%% seasonal pattern of activity depending on areas and ages

%%%%% seasonal pattern of mobility depending on areas and ages

%%%%% seasonal pattern of radius of gyration from home location depending on areas and ages

\section{Conclusion and discussion}
We have analyzed the social activity patterns of individuals in urban and rural areas of a country using the CDR data of a mobile phone service provider for the time span of one year. Using such a large data set, the average behaviour patterns of people, like the communication activity and mobility of each individual, can be deciphered on a daily, weekly, and seasonal basis. 

In our study the urban and rural areas are found to exhibit contrasting human behaviour patterns.
Using the population density determined from the ``home'' locations of individuals in conjunction with the weekly and seasonal patterns we were able to differentiate between three different types of areas: urban (Area 1-2), rural (Area 4-5), and intermediate (Area 3). This result is consistent with the fact that Areas 1 and 2 correspond to the central part of the metropolitan area, Area 3 to its periphery, and Areas 4 and 5 to the  sparsely populated areas further away from the metropolitan area. 
We find that individuals in densely populated areas, i.e. urban areas, communicate more and move less than in sparsely populated areas. This result matches with a study done in the United States, where the rural mobility was found to be relatively speaking higher than urban mobility \cite{pucher2004urban}. Such a behaviour can be attributed to the expansive nature of rural areas compared to urban ones, which results in homes and workplaces being dispersed over a broader expanse. Consequently, rural inhabitants are compelled to traverse greater distances on average than their urban counterparts. 

In addition we have found that on the basis of difference in seasonal patterns by area, the behaviour of people in this country are strongly bound by religious holidays, i.e., Easter and Christmas. During both holidays, individuals tend to stay at ``home'' but they communicate more actively during Christmas, while less on Easter. This influence of religious holidays tends to be more pronounced in rural areas than urban areas. On the other hand, the drop of communication activity and the peak in mobility in August coincides with the vacation time and appear more clearly in urban areas than in rural areas.  

Our analysis also finds that both the communication activity and mobility tend to be on average more frequent for females than for males, for all ages and areas. 
In addition, the gender difference in the communication activity in the urban areas is found to be larger than that in the rural areas. We also find that the gender difference is larger in teenage under 18 year old and middle age between 46 and 55 year old groups. 
The gender difference in the mobility is small but positive, i.e. in favour of females. Among people from 19 to 45 year old, males tend to move more actively than females do. 
Considering the mobility and radius of gyration from the ``home'' location, the senior people especially in the rural areas seem to move more around their ``home'' than younger age people \cite{pucher2004urban}.

As discussed above, the urban versus rural differences in human behaviour, focusing on the communication activity and mobility of individuals, can be successfully demonstrated. 
We expect that these features found by our analysis might be universal in countries having similar culture. Other countries with different cultural backgrounds would be interesting further targets of study. Finally, our present analysis does not reveal any structural features of human community for urban and rural areas, which could also be an interesting topic for further research.

\clearpage
\section{Materials and methods}
\subsection*{Data}
In this study, we analyze the human activity using anonymized mobile phone Call Detail Records (CDR) of the whole year 2008 from a mobile phone company in a southern European country with around 8 million service subscribers. 
From the CDR data, we use the following information for each call and SMS made between subscribers: the anonymized id of the subscriber who has received or made the call/ SMS, date and time of the call/ SMS, and the id of mobile phone tower that has been used for the call/ SMS. Each mobile phone tower is associated with a unique latitude and longitude information to identify its geographical location. 
The call record for each individual $i$ is then described as a time series of tower position $\bm{r}_i(t)$ used by the individual $i$ during the call/ SMS at time $t$. The position $\bm{r}_i(t)$ is represented by the geographical coordinates. Additionally, each user id is associated with their age and gender.

In this paper, we aim to reveal the pattern of usual human activity in a society by focusing on the socially relevant data, in which there is reciprocal communication between a pair of individuals. Thus we exclude the individuals who have only unilateral calls/ SMS to eliminate the activities which seem to be caused by systematic and external business. In order to discuss differences of individuals' activity patterns, we need to use the home position and therefore exclude individuals whose ``home'' position cannot be determined. 
After removing the unilateral calls/ SMS and the individuals without certain``home'' locations, the data set includes $5,950,640$ individuals in total. For our analysis about gender and age differences, we use a subset of the data which includes $748,770$ females and $1,060,589$ males.

\subsection*{Communication activity}
The communication activity $A_i$ of the individual $i$ is defined as the total number of incoming and outgoing voice calls along with SMS messages made by them in the time period of $\Delta T$ as  
\begin{equation}
A_i(\Delta T) = \sum_{t \in \Delta T} e_i(t),
\end{equation}
where for the individual $i$ making a call or SMS at time $t$ we record an activity event $e_i(t) = 1$ or otherwise $e_i(t) = 0$. 

\subsection*{Mobility}
The mobility $M_i$ of the individual $i$ is measured using the root mean square displacement, 
\begin{equation}
 M_i(\Delta T) = \sqrt{\frac{1}{A_i(\Delta T)}\sum_{t \in \Delta T}|\delta \bm{r}_{i}(t)|^2}, 
\end{equation}
 where the $|\delta \bm{r}_{i}(t)| = |\bm{r}_i(t)-\bm{r}_i(t+1)|$ represents the distance between two time records $t$ and $t+1$.

\subsection*{Radius of gyration from ``home'' location}
 As another metric for the mobility of individuals, we also calculate the radius of gyration from the ``home'' location of individual $i$ for a time period $\Delta T$ as follows, 
 \begin{equation}
 R_{g,i}(\Delta T) = \sqrt{\frac{1}{A_i(\Delta T)}\sum_{t \in \Delta T} |\delta \bm{r}_{i}^{\rm{home}}(t)|^2},
 \end{equation}
 where $|\delta \bm{r}_{i}^{\rm{home}}(t)| = |\bm{r}_i(t)-\bm{r}_{i}^{\rm{home}}|$ represents the distance from his or her ``home'' location $\bm{r}_{i}^{\rm{home}}$ at time $t$.

\subsection*{``Home'' location} 
The average pattern of daily activity of all egos has two peaks and both of them are well fitted by the normal distribution function $G(\mu, \sigma)$ where the day-time peak $\mu_{\rm{day}}$ appears to be at 12:45 pm ($\sigma_{\rm{day}}=2.72$) and the evening peak $\mu_{\rm{night}}$ at 7:38 pm ($\sigma_{\rm{night}}=3.14$) (see Fig.~\ref{sfig_daily_pattern}). 
During the night time low activity period, the individuals are assumed to stay at their home. We then define the ``home'' location $\bm{r}_i^{\rm{home}}$ of the individual $i$ as his or her average position during that time period, $\bm{r}_{i}^{\rm{home}}=\langle \bm{r}_{i}(t)\rangle_{\rm{rest}}$, where $\langle \cdots \rangle_{\rm{rest}}$ denotes the average over his or her location from 1 am to 7 am on a basis of the most inactive period. 

%%%%%%%%%%%%%%%%%%%%%%%%%%%%%%%%%%%%%%%%%%%%%%

\clearpage

\bibliography{bmc_article} 

\subsection*{Availability of data and materials}
The datasets generated during  and/ or analyzed during the current study are not publicly available due to a signed NDA but are available from the corresponding author on reasonable request.

\subsection*{Competing interests}
The authors declare that they have no competing interests.

\subsection*{Funding}
F.O. was supported by JST, PRESTO grant number JPMJPR2121, JAPAN. 
C.R. and K.K. acknowledge support from EU HORIZON 2020 INFRAIA-1-2014-2015 program project (SoBigData) No. 654024 and INFRAIA-2019-1 (SoBigData++) No. 871042 as well as NordForsk Programme for Interdisciplinary Research project “The Network Dynamics of Ethnic Integration”.

\subsection*{Author's contributions}
F.O. analyzed and interpreted the data, and was contribute to writing the manuscript. 
C.R. analyzed and interpreted the data, and was contribute to writing the manuscript.
K.K. supervised the project and interpreted the data, as well as contributed to writing the manuscript. All authors read and approved the final manuscript.

\subsection*{Acknowledgements}
F.O. thanks for the hospitality of Aalto University School of Science. 

\clearpage

\section*{Appendix}
\subsection*{Daily pattern of communication activity} 
\begin{figure}[h!]
\centering
\includegraphics[width=0.5\linewidth]{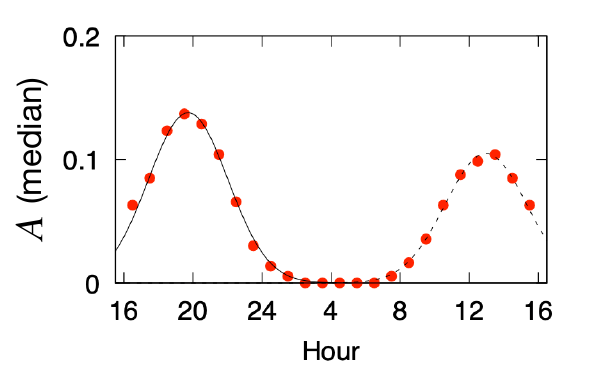}
\caption{Daily pattern of communication activity of all individuals.}
\label{sfig_daily_pattern}
\end{figure}

The daily pattern of communication activity shows a bimodal distribution as can be seen in Fig.~\ref{sfig_daily_pattern}. The two peaks are well fitted with Gaussian distribution functions $G(\mu, \sigma)$ where the day-time peak ($\mu_{night}=$12:58 pm, $\sigma_{night}=$2.36) and the evening peak ($\mu_{night}=$7:43 pm, $\sigma_{night}=$2.31). The low activity night time is defined as from $1$ am to $7$ am on a basis of the most inactive period.
%%%%% daily pattern communication activity with Gaussian fit

\subsection*{Daily, weekly, and seasonal patterns of communication activity and mobility} 
\begin{figure}[h]
\centering
\includegraphics[width=0.8\linewidth]{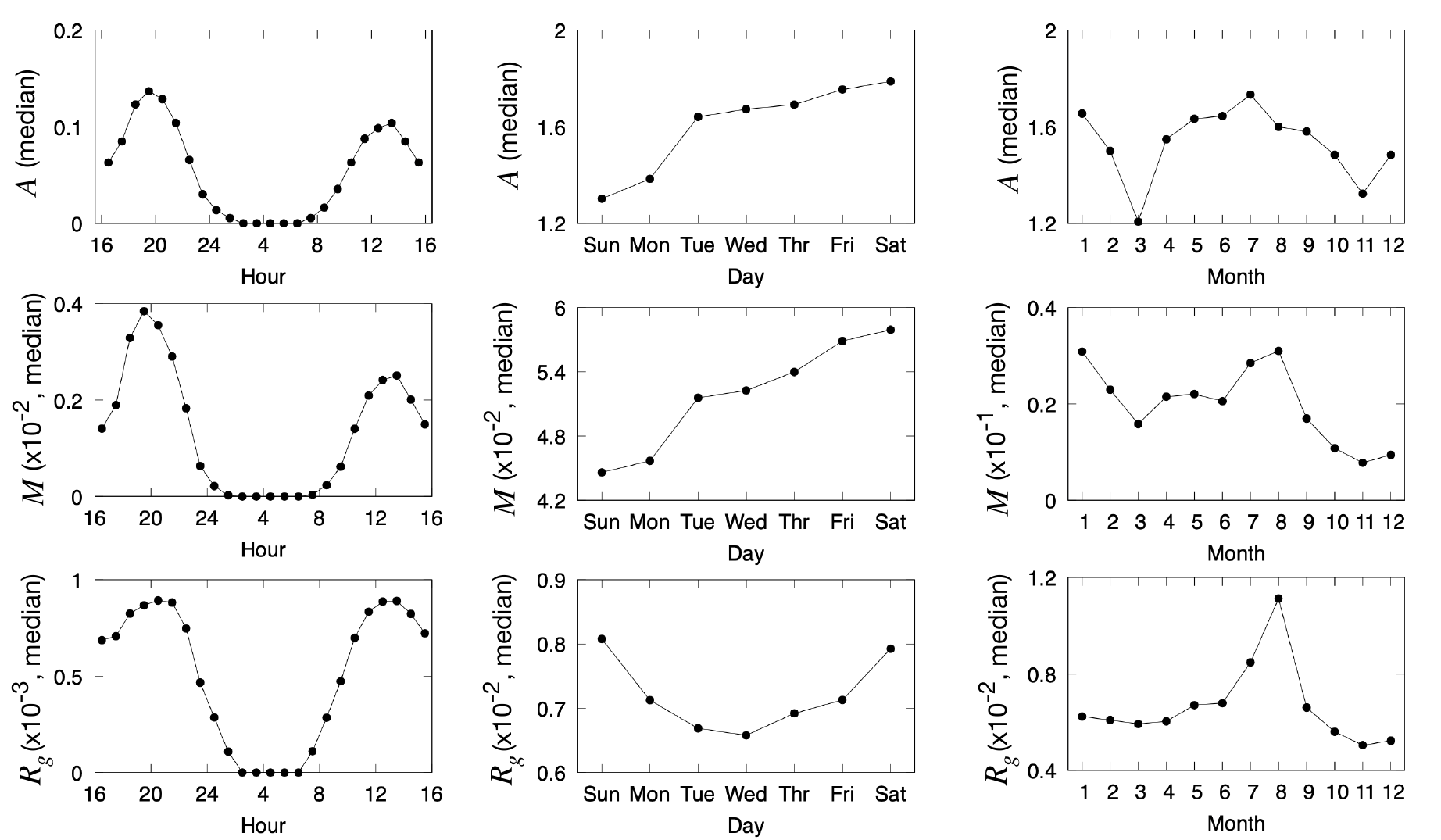}
  \caption{Average pattern of communication activity and mobility.Daily, weekly, and seasonal patterns of (upper) communication activity $A$ , (middle) mobility $M$, and (bottom) the radius of gyration from their home location $R_g$.}
  \label{sfig_average_pattern}
\end{figure}

The daily pattern of the communication activity $A$ and mobility $M$ (Rmsd) show a bimodal distribution plotted in Fig.~\ref{sfig_average_pattern}. The weekly patterns of $A$ and $M$ show that both values increase from Monday to Saturday and reach a minimum on Sunday, with more frequent communication and mobility during weekdays accompanied with staying at ``home'' on weekends.
The daily pattern of the radius of gyration from the ``home'' location $R_g$ also shows a bimodal distribution and the most inactive period of $R_g$ is consistent with the period from 1 am to 7 am. The weekly pattern of $R_g$ shows a minimum on Wednesdays. The seasonal pattern of $R_g$ reaches a peak in August and tends to be minimum in November and December. The $R_g$ seems to be slightly larger after January than from October to December.

%%%%% Daily, weekly, and seasonal patterns of all A, M, and Rg

\subsection*{Rank size distribution of the population density}
\begin{figure}[h]
\centering
\includegraphics[width=0.4\linewidth]{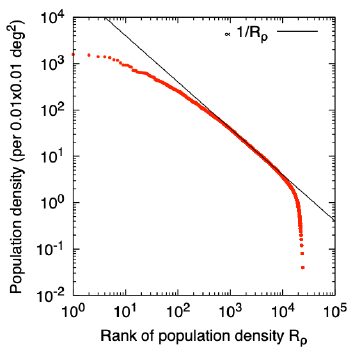}
  \caption{Rank size distribution of the population density.} 
  \label{sfig_drank}
\end{figure}

Average population density (see Fig.~\ref{sfig_drank}) (number density of egos per $0.01 \times 0.01$ deg$^{2}$ corresponding roughly an area of $1 \times 1$ km$^{2}$) is $1252.9$ for Area1, 418.7 for Area2, 83.2 for Area3, 10.0 for Area4, and 1.5 for Area5. The average density of Area3 is comparable to the actual value of population density of this country, $90$ people per km$^{2}$ in 2008.

%%%%% rank size plot of population density 

\subsection*{Weekly and seasonal patterns of communication activity and mobility by area}
\begin{figure}[h]
\centering
\includegraphics[width=0.5\linewidth]{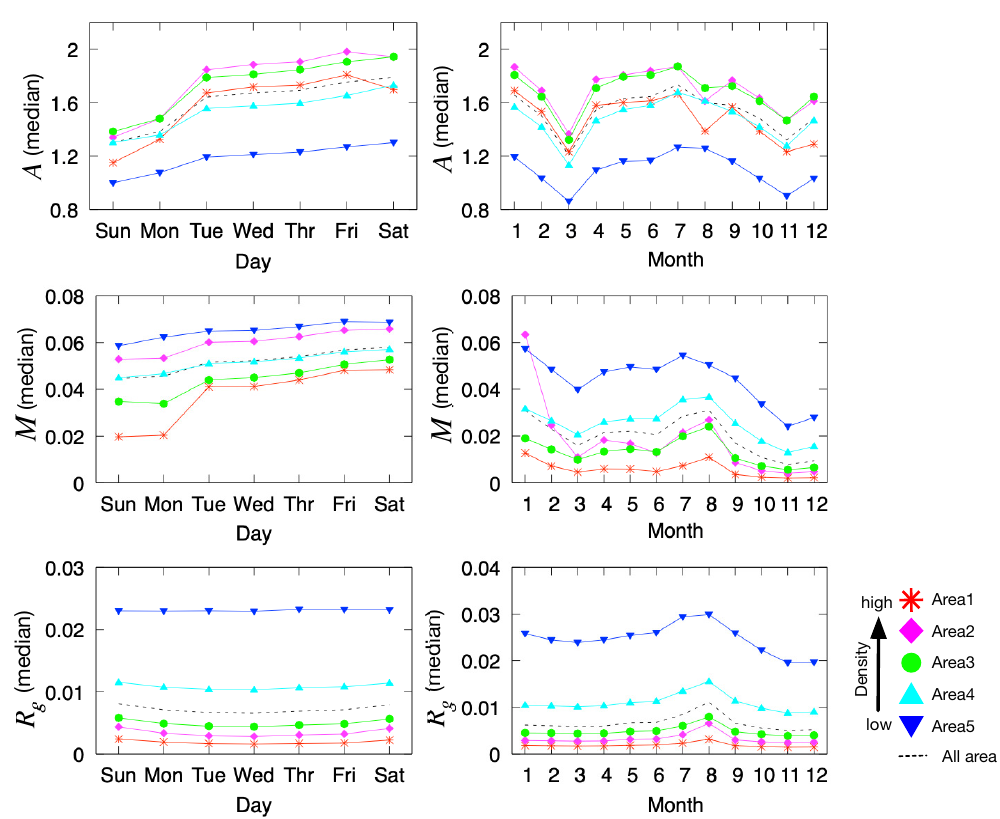}
  \caption{Weekly and seasonal patterns of communication activity ($A$), mobility ($M$, Rmsd), and the radius of gyration from their ``home'' location $R_g$ by areas (Area 1 to Area 5) in 2008.} 
    \label{sfig_AMpatterns_nominal}
\end{figure}

Weekly and seasonal patterns of communication activity $A$ show that it tends to be higher in the densely populated areas (Area 1, Area 2, and Area 3), where the communication activity shows a drop in August. On the other hand, weekly and seasonal patterns of mobility $M$ and the radius of gyration from the ``home'' location $R_g$ show that $M$ and $R_g$ increase as the population density decreases as shown in Fig.~\ref{sfig_AMpatterns_nominal}. 

%%%%% average pattern of activity and mobility

\subsection*{Area and age dependencies of communication activity and mobility though a year 2008}
\begin{figure}[h]
\centering
\includegraphics[width=0.5\linewidth]{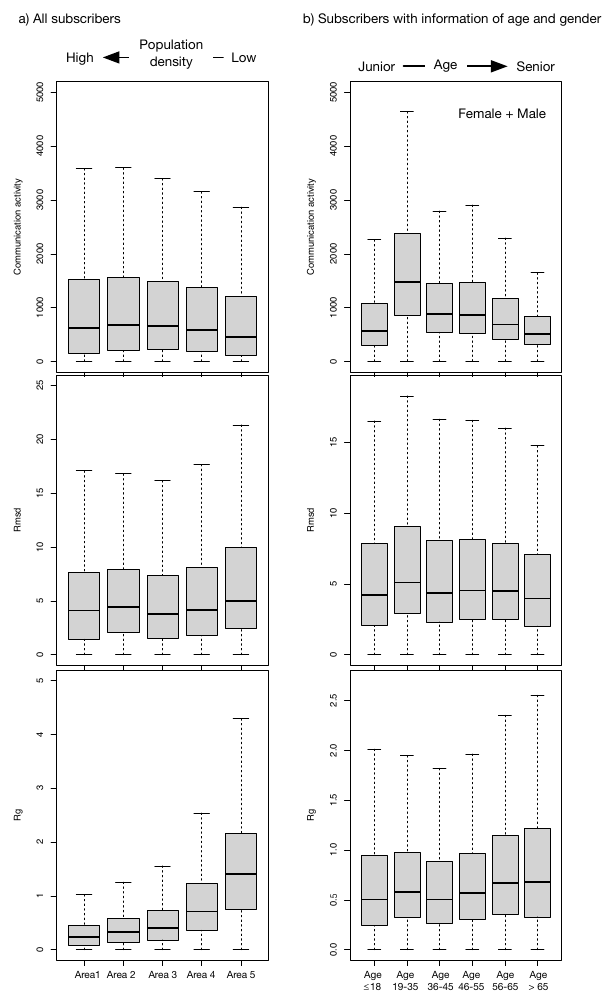}
  \caption{Area and age dependence of communication activity, mobility (Rmsd), and the radius of gyration from their ``home'' location in the year.}
 \label{sfig_totalAMRg}
\end{figure}

Communication activity $A$ tends to decrease slightly as the population density decreases. For early adulthood (between 19 and 35 year old) $A$ is highest and decreases for age groups after 36 year old. 
On the other hand, the mobility $M$ and the radius of gyration from the ``home'' location $R_g$ seem to increase with decreasing population density. As the age increase,$R_g$ seems to be slightly higher in the senior age group than in the younger age groups while $M$ tends to be more or less the same (see Fig.~\ref{sfig_totalAMRg}).

%%%%% boxplot of communication activity, rmsd, and rg by age group

\end{document}